\documentclass[aps,preprint]{revtex4}%
\usepackage{amsfonts}
\usepackage{amsmath}
\usepackage{amssymb}
\usepackage{graphicx}%
\begin{document}
\title{Large-$N$ droplets in two dimensions}
\author{Dean Lee}
\affiliation{Department of Physics, North Carolina State University, Raleigh, NC 27695}

\begin{abstract}
Using lattice effective field theory, we study the ground state binding energy
of $N$ distinct particles in two dimensions with equal mass interacting weakly
via an attractive $SU(N)$-symmetric short range potential. \ We find that in
the limit of zero range and large $N$, the ratio of binding energies
$B_{N}/B_{N-1}$ approaches the value $8.3(6).$

\end{abstract}
\maketitle

\section{Introduction}

We consider the ground state of $N$ distinct particles in two dimensions with
equal mass interacting weakly via an attractive $SU(N)$-symmetric short range
potential. \ Since the ground state is completely symmetric this is equivalent
to the problem of $N$ weakly-bound identical bosons. \ The self-bound
two-dimensional system with a realistic van der Waals potential is relevant to
the properties of adatoms on surfaces. \ In this work, however, we address the
question of what happens as the range of the interaction goes to zero,%
\begin{equation}
V(\vec{x}_{1},\cdots,\vec{x}_{N})\rightarrow C\sum_{1\leq i<j\leq N}%
\delta^{(2)}(\vec{x}_{i}-\vec{x}_{j}).
\end{equation}

Let $B_{N}$ be the ground state binding energy of the $N$-particle system in
the zero range limit. \ The first calculation of $B_{3}/B_{2}$ was given in
\cite{Bruch:1979}. \ The precision of this calculation was improved by
\cite{Nielsen:1999}, and most recently a precise value of $B_{3}%
/B_{2}=16.522688(1)$ was given in \cite{Hammer:2004x}$.$ \ There have also
been studies of the four- and five-particle systems
\cite{Tjon:1980,Lim:1980,Vranjes:2002}. \ But range corrections for these
studies appear to be very large, and the first precise determination of
$B_{4}/B_{2}$ in the zero range limit was only recently given in
\cite{Platter:2004x}, yielding a value of $B_{4}/B_{2}=197.3(1).$

The behavior of $B_{N}$ in the large-$N$ limit was also recently discussed in
\cite{Hammer:2004x}. \ They showed that due to the weakening of the attractive
coupling at short distance scales, the large-$N$ droplet system could be
treated classically. \ This yielded a prediction for the ratio of the binding
energies in the large-$N$ limit,%
\begin{equation}
\lim_{N\rightarrow\infty}\frac{B_{N}}{B_{N-1}}\simeq8.567\text{.}%
\end{equation}
In \cite{Blume:2004} the $N$-particle system for $N\leq7$ was investigated
using diffusion Monte Carlo with both a Lennard-Jones potential and a more
realistic helium-helium potential. \ However the results showed that range
corrections were too large to allow for a determination of $B_{N}/B_{N-1}$ for
large $N$.

Although there is no known system of atomic or molecular clusters that
displays the physics of the zero range limit for large $N$, the topic is
interesting for several reasons. \ With recent advances in laser trapping
techniques it is now possible to produce many-body quantum systems on a
two-dimensional optical lattice. \ Much of the attention has been devoted to
the Bose-Hubbard model with repulsive on-site interactions
\cite{Batrouni:2002,Scarola:2005}, but the weakly attractive $N$-boson system
can also be studied. \ In that case computational lattice studies such as this
would be of immediate relevance. \ Our system also raises interesting
questions about the convergence of effective field theory and the large-$N$
limit. \ Results of previous numerical studies suggest that it is surprisingly
difficult to reach the zero range and large-$N$ limits at the same time. \ We
explore why this is the case and what can be done to overcome some of the
difficulties. \ Similar issues arise in systems of higher-spin fermions in
optical traps and lattices. \ In these systems the competition between short
range interactions and large-$N$ effects can determine properties of the
ground state, two-particle pairing versus multi-particle clustering
\cite{Wu:2003a,Wu:2004a}.

In this paper we study the $N$-particle system using lattice effective field
theory. \ The organization of our paper is as follows. \ We first discuss the
renormalization of the interaction coefficient in the two-particle system.
\ We discuss renormalization in the continuum with a sharp momentum cutoff and
then on the lattice.\ \ After that we address two features of the large-$N$
limit. \ The first is a rescaling technique that cancels some of the nonzero
range corrections from the ratio $B_{N}/B_{N-1}$. \ The other is an
overlapping interaction problem that occurs when many particles lie within a
region the size of the range of the potential. \ We show that this problem can
produce large systematic errors that grow with $N$. \ The strength of the
overlapping interaction must be reduced if we wish to probe zero range physics
accurately for large $N$. \ We demonstrate one way of doing this which
exploits an unusual feature of the discrete Hubbard-Stratonovich
transformation \cite{Hirsch:1983}. \ Using lowest-order lattice effective
field theory, we compute $B_{N}/B_{N-1}$ for $N\leq10$. \ Extrapolating to the
limit $N\rightarrow\infty$, we find the result%
\begin{equation}
\lim_{N\rightarrow\infty}\frac{B_{N}}{B_{N-1}}=8.3(6)\text{.}%
\end{equation}

\section{Two-particle system and renormalized coupling}

\begin{figure}
[ptb]
\begin{center}
\includegraphics[
height=0.8856in,
width=2.284in
]%
{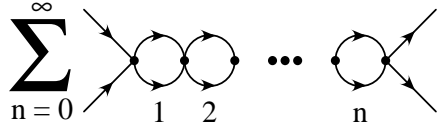}%
\caption{Diagrams contributing to two-particle scattering for a zero-range
potential.}%
\label{scattering}%
\end{center}
\end{figure}

We begin by reviewing the two-particle system in the continuum formalism with
a sharp cutoff, $\Lambda$, on the magnitude of the momentum. \ For a zero
range potential,%
\begin{equation}
V(\vec{x}_{1},\cdots,\vec{x}_{N})=C\sum_{1\leq i<j\leq N}\delta^{(2)}(\vec
{x}_{i}-\vec{x}_{j}),
\end{equation}
the diagrams which contribute to two-particle scattering are shown in Fig.
\ref{scattering}.%
We let $m$ be the particle mass. \ In order that the bound state pole in the
rest frame occurs at energy $E=-B_{\text{2}}$, we get the constraint%
\begin{equation}
-\frac{1}{C}=\frac{1}{2\pi}\int_{0}^{\Lambda}\frac{pdp}{B_{\text{2}}%
+\frac{p^{2}}{m}}=\frac{m}{4\pi}\ln\left(  \frac{mB_{2}+\Lambda^{2}}{mB_{2}%
}\right)  ,
\end{equation}
We can rewrite this as%
\begin{equation}
\frac{mB_{2}\Lambda^{-2}}{1+mB_{2}\Lambda^{-2}}=\exp\left[  \frac{4\pi}%
{Cm}\right]  ,
\end{equation}
and the bound state energy is given by%
\begin{equation}
mB_{2}\Lambda^{-2}=\exp\left[  \frac{4\pi}{Cm}\right]  +O\left[  \left(
mB_{2}\Lambda^{-2}\right)  ^{2}\right]  .
\end{equation}

We now consider the same calculation on the lattice. \ Let $a$ be the spatial
lattice spacing and $a_{t}$ be the temporal lattice spacing. We start with the
Hamiltonian lattice formulation where $a_{t}=0$. \ The standard lattice
Hamiltonian with nearest neighbor hopping has the form%
\begin{align}
H  &  =\frac{1}{2ma^{2}}\sum_{\vec{n}}\sum_{1\leq i\leq N}\sum_{l=x,y}\left[
2b_{i}^{\dagger}(\vec{n})b_{i}(\vec{n})-b_{i}^{\dagger}(\vec{n})b_{i}(\vec
{n}+\hat{l})-b_{i}^{\dagger}(\vec{n})b_{i}(\vec{n}-\hat{l})\right] \nonumber\\
&  +Ca^{-2}\sum_{\vec{n}}\sum_{1\leq i<j\leq N}b_{i}^{\dagger}(\vec{n}%
)b_{i}(\vec{n})b_{j}^{\dagger}(\vec{n})b_{j}(\vec{n}).
\end{align}
Here, $b_{i}(\vec{n})$ is an annihilation operator for a particle with flavor
$i$ at the spatial lattice site $\vec{n}$. \ The condition on $C$ in the
Hamiltonian lattice formalism is%
\begin{equation}
-\frac{1}{C}=\lim_{L\rightarrow\infty}\frac{1}{a^{2}L^{2}}\sum_{\vec{k}\text{
}\operatorname{integer}}\frac{1}{B_{\text{2}}+2\Omega_{_{\vec{k}}}},
\end{equation}
where $\Omega_{_{\vec{k}}}$ is the lattice kinetic energy and $L$ is the
length of the periodic lattice cube in lattice units. \ For the standard
lattice action
\begin{equation}
\Omega_{_{\vec{k}}}=\frac{1}{ma^{2}}\sum_{s=x,y}\left[  1-\cos\tfrac{2\pi
k_{s}}{L}\right]  . \label{standard}%
\end{equation}
For later reference we define $\omega$ as the momentum independent term inside
the summation in (\ref{standard}). \ So for the standard lattice action
$\omega=1$. \ We define the lattice cutoff momentum $\Lambda=\pi a^{-1}$.
\ Then in the limit $\Lambda\rightarrow\infty$,%
\begin{equation}
mB_{2}\Lambda^{-2}=B\exp\left[  \frac{4\pi}{Cm}\right]  +O\left[  \left(
mB_{2}\Lambda^{-2}\right)  ^{2}\right]  , \label{asymptotic}%
\end{equation}
where for the standard action $B\simeq3.24$.

In order to test the cutoff dependence of our lattice results, we also
consider actions with $O(a^{2})$-improved and $O(a^{4})$-improved kinetic
energies. \ $O(a^{2})$ and $O(a^{4})$ corrections to the interaction are not
included since these would entail a significant number of new interactions.
\ Because we are not performing a full $O(a^{2})$ or $O(a^{4})$ improvement,
we do not expect the improved kinetic energy actions to give qualitatively
better results than the standard action. \ However a comparison of the
different actions provides an additional check that the results reproduce
continuum limit behavior rather than lattice-dependent artifacts. \ For the
$O(a^{2})$-improved action the lattice kinetic energy is%
\begin{equation}
\Omega_{_{\vec{k}}}=\frac{1}{ma^{2}}\sum_{s=x,y}\left[  \tfrac{5}{4}-\tfrac
{4}{3}\cos\tfrac{2\pi k_{s}}{L}+\tfrac{1}{12}\cos\tfrac{4\pi k_{s}}{L}\right]
.
\end{equation}
In this case $\omega=\tfrac{5}{4}$ and $B\simeq1.79$, where $B$ is defined in
the asymptotic expression (\ref{asymptotic}). \ For the $O(a^{4})$-improved
action%
\begin{equation}
\Omega_{_{\vec{k}}}=\frac{1}{ma^{2}}\sum_{s=x,y}\left[  \tfrac{49}{36}%
-\tfrac{3}{2}\cos\tfrac{2\pi k_{s}}{L}+\tfrac{3}{20}\cos\tfrac{4\pi k_{s}}%
{L}-\tfrac{1}{90}\cos\tfrac{6\pi k_{s}}{L}\right]  ,
\end{equation}
$\omega=\tfrac{49}{36}$, and $B\simeq1.54$. \ As we increase the order of
improvement, $\Omega_{_{\vec{k}}}$ more closely approximates the continuum
kinetic energy and $B$ approaches the continuum sharp cutoff value of $1$. \ 

At nonzero temporal lattice spacing the same diagrams in Fig. \ref{scattering}
contribute to two-particle scattering. \ A derivation of the Feynman rules at
nonzero temporal lattice spacing for the analogous three-dimensional system
can be found in \cite{Lee:2004qd}, as well as a derivation of the bound state
pole condition. \ In two dimensions the strength of the interaction is given
by the transfer matrix element%
\begin{equation}
\left(  e^{-a_{t}Ca^{-2}}-1\right)  \left(  1-\omega\frac{2a_{t}}{ma^{2}%
}\right)  ^{2},
\end{equation}
while the free lattice propagator has the form%
\begin{equation}
\frac{1}{e^{-\tfrac{2\pi i}{L_{t}}k_{0}}-1+a_{t}\Omega_{_{\vec{k}}}}.
\end{equation}
$L_{t}$ is the total number of temporal lattice units and $k_{0}$ is an
integer from $0$ to $L_{t}-1$.

As we take $L_{t}\rightarrow\infty$, the energy in physical units becomes a
continuous variable. \ Requiring that the bound state pole in the rest frame
occurs at energy $E=-B_{\text{2}}$, we get the constraint%
\begin{equation}
\frac{1}{\left(  1-\omega\frac{2a_{t}}{ma^{2}}\right)  ^{2}\left(
e^{-a_{t}Ca^{-2}}-1\right)  }=\lim_{L\rightarrow\infty}\frac{1}{L^{2}}%
\sum_{\vec{k}\text{ integer}}\frac{1}{e^{a_{t}B_{2}}-1+2a_{t}\Omega_{_{\vec
{k}}}-a_{t}^{2}\Omega_{_{\vec{k}}}^{2}}. \label{temporal}%
\end{equation}
\ At nonzero temporal lattice spacing we therefore have%
\begin{equation}
mB_{2}\Lambda^{-2}=B(a_{t}m^{-1}a^{-2})\exp\left[  \frac{4\pi}{C^{\prime}%
m}\right]  +O\left[  \left(  mB_{2}\Lambda^{-2}\right)  ^{2}\right]  ,
\label{temporalasymptotic}%
\end{equation}
where%
\begin{equation}
C^{\prime}\equiv\frac{a^{2}}{a_{t}}\left(  1-\omega\frac{2a_{t}}{ma^{2}%
}\right)  ^{2}\left(  1-e^{-a_{t}Ca^{-2}}\right)  .
\end{equation}
In this case $B$ is a function of $a_{t}m^{-1}a^{-2}$ and is different for the
standard and improved lattice actions. \ For given values of $B_{2},a$, and
$a_{t}$, we determine $C$ in the infinite volume limit $L\rightarrow\infty$.
\ For $mB_{2}\Lambda^{-2}$ not too small, roughly $10^{-6}$ or larger, we use
the exact expression (\ref{temporal}) for sufficiently large values for $L$.
\ For smaller values of $mB_{2}\Lambda^{-2}$ it is more convenient to use the
asymptotic expression (\ref{temporalasymptotic}). \ But once the interaction
coefficient $C$ is determined and we proceed to the $N$-body system, such
extremely large lattice volumes are unnecessary. \ It suffices to consider
lattice systems larger than the characteristic size of the $N$-body droplet.
\ For large $N$ this is many orders of magnitude smaller than the
characteristic size of the two-body droplet.

\section{Ratios in the large-$N$ limit}

It has been suggested that the large-$N$ ground state wavefunction can be
described as a classical distribution \cite{Hammer:2004x}. \ If $R_{N}$ is the
characteristic size of the droplet, the distribution is proportional to
$\psi(r/R_{N})$ for some function $\psi$ independent of $N$, and the binding
energy $B_{N}$ is proportional $m^{-1}R_{N}^{-2}$. \ In order to determine
$\psi$, one integrates out high energy modes to determine the effective
coupling at energy $B_{N}$. \ If this picture of the large-$N$ droplet is
correct, then errors due to the finite cutoff momentum $\Lambda$ appear only
in the combination $mB_{N}\Lambda^{-2}$. \ Therefore if we measure binding
energies while keeping $mB_{N}\Lambda^{-2}$ fixed, much of the error cancels
in the ratio $B_{N}/B_{N-1}$. \ In essence we are using large-$N$ similarity
under rescaling to eliminate cutoff errors. \ If the classical droplet picture
is incorrect, then this technique will probably not reduce errors. \ The issue
will be settled when we analyze results of the Monte Carlo simulations.

Let $B_{N}(\Lambda)$ be the measured binding energy of the $N$-particle ground
state at cutoff momentum $\Lambda$. \ Conceptually it is simplest to regard
$m$ and $B_{2}$ as fixed quantities while we vary $\Lambda$. \ In the
continuum limit%
\begin{equation}
\lim_{\Lambda\rightarrow\infty}B_{N}(\Lambda)=B_{N}.
\end{equation}
Let $z>0$ be a parameter that measures proximity to the continuum limit,%
\begin{equation}
z=mB_{N}(\Lambda)\cdot\Lambda^{-2}.
\end{equation}
For a given $z$, we define the cutoff momentum $\Lambda(z,N)$ implicitly so
that
\begin{equation}
mB_{N}(\Lambda(z,N))\cdot(\Lambda(z,N))^{-2}=z.
\end{equation}
We define $f(z)$ as%
\begin{equation}
f(z)=\lim_{N\rightarrow\infty}\frac{1}{N}\ln\left[  B_{N}(\Lambda
(z,N))/B_{N}\right]  . \label{fz}%
\end{equation}
$f(z)$ measures the exponential growth of finite cutoff errors with increasing
$N$. \ We have%
\begin{equation}
\lim_{N\rightarrow\infty}\left\{  \ln\left[  \frac{B_{N}(\Lambda(z,N))}{B_{N}%
}\right]  -\ln\left[  \frac{B_{N-1}(\Lambda(z,N-1))}{B_{N-1}}\right]
\right\}  =f(z),
\end{equation}
and so%
\begin{equation}
\lim_{N\rightarrow\infty}\frac{B_{N}}{B_{N-1}}=e^{-f(z)}\lim_{N\rightarrow
\infty}\frac{B_{N}(\Lambda(z,N))}{B_{N-1}(\Lambda(z,N-1))}. \label{largeN}%
\end{equation}
Therefore so long as $\left\vert f(z)\right\vert \ll1,$ the large-$N$ ratio of
binding energies can be measured reliably. \ Other cutoff errors which do not
grow linearly with $N$ will cancel in the ratio%
\begin{equation}
\frac{B_{N}(\Lambda(z,N))}{B_{N-1}(\Lambda(z,N-1))}%
\end{equation}
as we take $N\rightarrow\infty$.

In our Monte Carlo lattice simulations it is more convenient to regard $m$ and
$\Lambda$ as fixed quantities while varying $B_{2}$. \ We define $B_{2}(z,N)$
implicitly by%
\begin{equation}
mB_{N}(B_{2}(z,N))\cdot\Lambda^{-2}=z.
\end{equation}
We are changing the overall physical scale when we change $B_{2}$, and so we
work with ratios $B_{N}/B_{2}$. \ The analog of the result (\ref{largeN}) is%
\begin{equation}
\lim_{N\rightarrow\infty}\frac{B_{N}}{B_{N-1}}=e^{-f(z)}\lim_{N\rightarrow
\infty}\frac{B_{N}(B_{2}(z,N))/B_{2}(z,N)}{B_{N-1}(B_{2}(z,N-1))/B_{2}%
(z,N-1)}.
\end{equation}

\section{Overlapping range and implicit $N$-body interaction}

Large range corrections can occur when many particles lie within a region the
size of the range of the potential, $\Lambda^{-1}$. \ The problem is most
severe when all $N$ particles lie in this localized region, and the potential
energy is amplified by a factor of $N(N-1)/2$. \ For a continuum potential
with a repulsive core, the result is a deep hole at the center of the
multiparticle wavefunction and a tendency towards underbinding or unbinding.
\ At lowest order in lattice effective field theory the effect goes in the
opposite direction. \ A spike forms at the center of the wavefunction when all
particles lie on the same lattice site, and the binding energy is too large.

Consider the state with $N$ particles at the same lattice site in the
Hamiltonian lattice formalism,
\begin{equation}
\left\vert \Pi^{N}\right\rangle =b_{1}^{\dagger}(\vec{n})b_{2}^{\dagger}%
(\vec{n})\cdots b_{N}^{\dagger}(\vec{n})\left\vert 0\right\rangle .
\end{equation}
The expectation value of the potential energy for this localized state is%
\begin{equation}
\left\langle \Pi^{N}\right\vert V\left\vert \Pi^{N}\right\rangle
=\frac{CN(N-1)}{2a^{2}}.
\end{equation}
This potential energy can be regarded as an implicit $N$-body contact
interaction produced\ by overlapping two-body interactions. $\ $In the
continuum limit we know that the importance of this $N$-body contact
interaction is suppressed by many powers of the small parameter $z=mB_{N}%
\Lambda^{-2}$. \ However the situation at finite $\Lambda$ can be quite
different from the continuum limit if the potential energy per particle for
the localized state $\left\vert \Pi^{N}\right\rangle $ is as large as the
cutoff energy scale,%
\begin{equation}
\left\vert \frac{C(N-1)}{2a^{2}}\right\vert \gtrsim\frac{\pi^{2}}{ma^{2}}.
\end{equation}

To lowest order in $mB_{2}\Lambda^{-2}$, the renormalized coupling is%
\begin{align}
C  &  =\frac{4\pi}{m\ln\left(  mB_{2}\Lambda^{-2}\right)  }\nonumber\\
&  =\frac{4\pi}{m\ln\left(  mB_{N}\Lambda^{-2}\right)  -m\ln\left(
B_{N}/B_{2}\right)  }.
\end{align}
For large $N$%
\begin{equation}
C\simeq\frac{4\pi}{m\ln z-mN\ln\beta},
\end{equation}
where
\begin{equation}
\beta=\lim_{N\rightarrow\infty}\frac{B_{N}}{B_{N-1}}.
\end{equation}
Then%
\begin{equation}
-\frac{C(N-1)}{2a^{2}}\simeq\frac{\pi^{2}}{ma^{2}}\left[  \frac{2\pi^{-1}}%
{\ln\beta-\frac{1}{N}\ln z}\right]  .
\end{equation}
In the continuum limit the problem goes away since%
\begin{equation}
\frac{1}{\ln\beta-\frac{1}{N}\ln z}\rightarrow0.
\end{equation}
However the convergence is slow and requires $z\ll e^{-N}$. \ For actual
lattice simulations it is therefore necessary to limit the size of the
implicit $N$-body contact interaction.

\section{Discrete Hubbard-Stratonovich transformation}

There are several ways to deal with the large implicit $N$-body contact
interaction. \ On the lattice there is one method which is particularly
convenient. \ This is to write the two-body interaction using a discrete
Hubbard-Stratonovich transformation \cite{Hirsch:1983}$.$ \ The discrete
Hubbard-Stratonovich reproduces the two-body contact interaction exactly.
\ Typically it is used for systems with spin-$1/2$ fermions where Pauli
exclusion implies that there are no $N$-body contact interactions beyond
$N=2$. \ It seems therefore that the properties of the transformation for
$N\geq3$ has not been discussed in the literature. \ In the following we show
that when a discrete Hubbard-Stratonovich transformation is used, the temporal
lattice spacing regulates the strength of the implicit $N$-body contact interaction.

For simplicity we show only the interaction part of the Hamiltonian. \ The
exponential of the two-body interaction at site $\vec{n}$ over a Euclidean
time step $a_{t}$ is%
\begin{equation}
e^{-a_{t}H_{\text{int}}}=\exp\left[  -a_{t}Ca^{-2}\sum_{1\leq i<j\leq N}%
b_{i}^{\dagger}(\vec{n})b_{i}(\vec{n})b_{j}^{\dagger}(\vec{n})b_{j}(\vec
{n})\right]  .
\end{equation}
The discrete Hubbard-Stratonovich transformation amounts to making the
replacement%
\begin{equation}
e^{-a_{t}H_{\text{int}}}\rightarrow\frac{1}{2}\sum_{s(\vec{n})=\pm1}%
\exp\left[  -\left(  \frac{1}{2}a_{t}Ca^{-2}+\lambda s(\vec{n})\right)
\left(  \sum_{1\leq i\leq N}b_{i}^{\dag}(\vec{n})b_{i}(\vec{n})-1\right)
\right]  ,
\end{equation}
where%
\begin{equation}
\cosh\lambda=\exp\left(  -\frac{1}{2}a_{t}Ca^{-2}\right)  ,\qquad\lambda\geq0.
\end{equation}

To see that this has all the desired properties, let us define%
\begin{equation}
A(K)=\frac{1}{2}\sum_{s(\vec{n})=\pm1}\exp\left[  -\left(  \frac{1}{2}%
a_{t}Ca^{-2}+\lambda s(\vec{n})\right)  (K-1)\right]  ,
\end{equation}
for nonnegative integer $K$. \ We note that $A(0)=A(1)=1$, and $A(2)=\exp
\left(  -a_{t}Ca^{-2}\right)  $. \ These are precisely the expectation values
of $e^{-a_{t}H_{\text{int}}}$ for $K=0,1,2$ distinct particles at lattice site
$\vec{n}$. \ When $K\geq3$ but $\lambda(K-1)\ll1$, we find%
\begin{equation}
A(K)\simeq\exp\left[  -a_{t}Ca^{-2}\frac{K(K-1)}{2}\right]  .
\end{equation}
This is also the expectation value of $e^{-a_{t}H_{\text{int}}}$ for $K$
distinct particles at lattice site $\vec{n}$. \ However when $K\geq3$ and
$\lambda(K-1)$ $\gg1$,%
\begin{equation}
A(K)\simeq\frac{1}{2}\exp\left[  \left(  -\frac{1}{2}a_{t}Ca^{-2}%
+\lambda\right)  (K-1)\right]  .
\end{equation}
The total potential energy of the $K$-particle localized state, $\left\vert
\Pi^{K}\right\rangle $, no longer increases quadratically with $K$. \ The
temporal lattice spacing $a_{t}$ acts as an auxiliary ultraviolet regulator
that limits the size of the implicit $K$-body contact interaction. $\ $When
$K\leq2$ or the implicit $K$-body contact interaction is smaller than
$a_{t}^{-1}$, we have the unaltered result,%
\begin{equation}
\left\langle \Pi^{K}\right\vert V\left\vert \Pi^{K}\right\rangle \simeq
\frac{CK(K-1)}{2a^{2}}.
\end{equation}
When $K>2$ and the implicit $K$-body contact interaction exceeds $a_{t}^{-1}$,
then the regulator takes effect and we have%
\begin{equation}
\left\langle \Pi^{K}\right\vert V\left\vert \Pi^{K}\right\rangle \simeq
a_{t}^{-1}\left[  \left(  \frac{1}{2}a_{t}Ca^{-2}-\lambda\right)
(K-1)+\ln2\right]  .
\end{equation}

\section{Algorithm}

The standard lattice action we use for our simulations is%
\begin{align}
&  \sum_{\vec{n},n_{t},i}\left[  c_{i}^{\ast}(\vec{n},n_{t})c_{i}(\vec
{n},n_{t}+1)-e^{-\tfrac{a_{t}Ca^{-2}}{2}-\lambda s(\vec{n},n_{t})}\left(
1-\frac{2a_{t}}{ma^{2}}\right)  c_{i}^{\ast}(\vec{n},n_{t})c_{i}(\vec{n}%
,n_{t})\right] \nonumber\\
&  -\frac{a_{t}}{2ma^{2}}\sum_{\vec{n},n_{t},l,i}\left[  c_{i}^{\ast}(\vec
{n},n_{t})c_{i}(\vec{n}+\hat{l},n_{t})+c_{i}^{\ast}(\vec{n},n_{t})c_{i}%
(\vec{n}-\hat{l},n_{t})\right]  -\sum_{\vec{n},n_{t}}\lambda s(\vec{n},n_{t}),
\end{align}
where $n_{t}$ is the temporal lattice coordinate, $c_{i}$ is the path
integration field for the particle of type $i$, and $s$ is the discrete
Hubbard-Stratonovich field which takes values $\pm1$. \ We have used the
lattice conventions developed in \cite{Lee:2004si, Lee:2004qd} for a
three-dimensional lattice. \ The choice of Bose/Fermi statistics for $c_{i}$
is irrelevant since we consider systems with no more than one particle of each type.

In order to compute the ground state binding energy $B_{N}$ we consider the
correlation function%
\begin{equation}
Z_{N}(t)=\left\langle \Psi_{N}^{0}\right\vert e^{-Ht}\left\vert \Psi_{N}%
^{0}\right\rangle ,
\end{equation}
where the initial/final state is the state with all $N$ particles at zero
momentum,%
\begin{equation}
\left\vert \Psi_{N}^{0}\right\rangle =\tilde{b}_{1}^{\dag}(0)\tilde{b}%
_{2}^{\dag}(0)\cdots\tilde{b}_{N}^{\dag}(0)\left\vert 0\right\rangle .
\end{equation}
$\left\vert \Psi_{N}^{0}\right\rangle $ is also the ground state of the
non-interacting system. \ We refer to $t$ as Euclidean time and define%
\begin{equation}
E_{N}(t)=-\frac{\partial}{\partial t}\left[  \ln Z_{N}(t)\right]  .
\end{equation}
Then as $t\rightarrow+\infty$, $E_{N}(t)$ converges to $-B_{N}$, the ground
state energy of the interacting $N$-particle system. \ The only assumption is
that the ground state has a nonvanishing overlap with the ground state of the
non-interacting system.

The conversion of the lattice action to a transfer matrix formalism at fixed
particle number has been discussed in \cite{Borasoy:2005yc}. \ We use the same
transfer matrix derived there, except in this case we keep the discrete
Hubbard-Stratonovich field and calculate the sum over configurations,%
\begin{equation}
Z_{N}(t)\varpropto\sum_{s}e^{-\sum_{\vec{n},n_{t}}\lambda s(\vec{n},n_{t}%
)}\left\langle \Psi_{N}^{0}\right\vert T\left[  e^{-H(s)t}\right]  \left\vert
\Psi_{N}^{0}\right\rangle ,
\end{equation}
$H(s)$ consists of only single-body operators interacting with the background
Hubbard-Stratonovich field. \ We can write the full $N$-particle matrix
element as the $N^{\text{th}}$ power of the single-particle matrix element,%
\begin{equation}
\left\langle \Psi_{N}^{0}\right\vert T\left[  e^{-H(s)t}\right]  \left\vert
\Psi_{N}^{0}\right\rangle \propto\left[  M(s,t)\right]  ^{N},
\end{equation}%
\begin{equation}
M(s,t)=\left\langle \vec{k}=0\right\vert T\left[  e^{-H(s)t}\right]
\left\vert \vec{k}=0\right\rangle ,
\end{equation}
where $\left\vert \vec{k}=0\right\rangle $ is a single-particle state with
zero momentum. \ Our time-ordered exponential notation, $T\left[
e^{-H(s)t}\right]  $, is shorthand for the time-ordered product of single-body
transfer matrices at each time step,%
\begin{equation}
T\left[  e^{-H(s)t}\right]  =M_{(L_{t}-1)}\cdot\ldots\cdot M_{(n_{t})}%
\cdot\ldots\cdot M_{(1)}\cdot M_{(0)},
\end{equation}
where $L_{t}$ is the total number of lattice time steps and $t=L_{t}a_{t}$.
\ If the particle stays at the same spatial lattice site from time step
$n_{t}$ to $n_{t}+1$, then the corresponding matrix element of $M_{(n_{t})}$
is%
\begin{equation}
e^{-\tfrac{a_{t}Ca^{-2}}{2}-\lambda s(\vec{n},n_{t})}\left(  1-\frac{2a_{t}%
}{ma^{2}}\right)  .
\end{equation}
If the particle hops to a neighboring lattice site from time step $n_{t}$ to
$n_{t}+1$ then the corresponding matrix element of $M_{(n_{t})}$ is
$\frac{a_{t}}{2ma^{2}}$. \ All other elements of $M_{(n_{t})}$ are zero.

We sample configurations according to the weight%
\begin{equation}
\exp\left\{  \sum_{\vec{n},n_{t}}\lambda s(\vec{n},n_{t})+N\log\left[
M(s,t_{\text{end}})\right]  \right\}  ,
\end{equation}
where $t_{\text{end}}$ is the largest Euclidean time at which we wish to
measure $Z_{N}(t)$. \ We use a simple heat bath/Metropolis update procedure.
\ For each configuration the observable that we compute is%
\begin{equation}
O(s,t)=\frac{\left[  M(s,t)\right]  ^{N}}{\left[  M(s,t_{\text{end}})\right]
^{N}},
\end{equation}
for $t<t_{\text{end}}$. \ This is the same general technique that was used in
\cite{Lee:2005fk}. \ By taking the ensemble average of $O(s,t)$ we are able to
calculate%
\begin{equation}
\frac{Z_{N}(t)}{Z_{N}(t_{\text{end}})}.
\end{equation}

\section{Results}

For each simulation we have computed roughly $2\times10^{5}$ successful heat
bath/Metropolis updates for each lattice site, split across four processors
running completely independent trajectories. \ Averages and errors were
calculated by comparing the results of each processor. \ The codes were based
on existing codes used for light nuclei in three-dimensions and we have kept
some of the same input parameters relevant for the light nuclei system. \ We
use a mass of $m=$ $939$ MeV and keep the spatial lattice spacing fixed at
$a=(40$ MeV$)^{-1}$. \ This corresponds with $\Lambda=\pi a^{-1}\simeq126$ MeV
and cutoff energy $\Lambda^{2}/m=16.8$ MeV. \ Clearly these input parameters
in raw form are not appropriate for atomic clusters. \ Therefore we translate
of all of the parameters in terms of dimensionless numbers which can then be
easily applied to any two-dimensional droplet system.

We have already defined the dimensionless ratio $z,$%
\begin{equation}
z=\frac{B_{N}}{\Lambda^{2}/m}=B_{N}ma^{2}\pi^{-2}.
\end{equation}
$z$ measures the ratio of $B_{N}$ to the cutoff energy. In most cases it is
clear which $N$ we are referring to and so we use the simple notation $z$.
\ When there is some possibility of confusion we include the $N$ subscript,
$z_{N}$.

We also define $\varepsilon$,
\begin{equation}
\varepsilon=\frac{\pi a_{t}^{-1}}{\Lambda^{2}/m}=a_{t}^{-1}ma^{2}\pi^{-1}.
\end{equation}
A small value for $\varepsilon$ indicates that there is a significant amount
of high frequency regularization provided by the nonzero temporal lattice
spacing $a_{t}$. \ A large value for $\varepsilon$ means that we are close to
the Hamiltonian limit, $a_{t}\rightarrow0$. \ There is little or no
regularization of high frequency modes and most of the regularization is
provided by the momentum cutoff $\Lambda$.

We adjust the two-particle binding energy $B_{2}$ in order to study the finite
cutoff dependence. \ Since we keep $\Lambda$ fixed, our value $B_{2}$ will
decrease as go to larger values of $N$. \ For convenience we use the shorthand%
\begin{equation}
b_{N}=B_{N}/B_{2}%
\end{equation}
for the dimensionless ratio of the binding energies. \ For each data point we
increase the spatial length and temporal extent of the lattice until the
finite volume/time errors are clearly smaller than the statistical errors.
\ The largest lattice system we simulate is $9\times9\times260.$

\begin{figure}
[ptb]
\begin{center}
\includegraphics[
height=4.2281in,
width=3.9453in
]%
{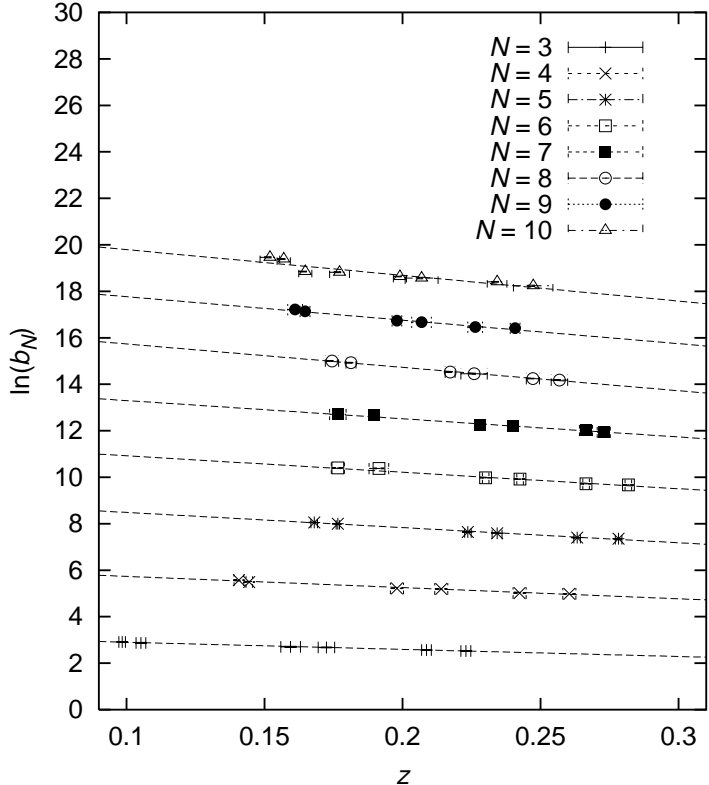}%
\caption{$\ln(b_{N})$ versus $z$. \ We use the $O(a^{4})$-improved action and
$\varepsilon=3.7$.}%
\label{z}%
\end{center}
\end{figure}

We have computed $b_{N}$ for $N\leq10$ for a wide range of values for $B_{2}$
using the $O(a^{4})$-improved action and $a_{t}=(20$ MeV$)^{-1}$, which
corresponds with $\varepsilon=3.7$. \ The results are shown as a plot of
$\ln(b_{N})$ versus $z$ in Fig. \ref{z}. \ We see that there is considerable
dependence on $z$. \ The dependence appears to be roughly linear in $z$ for
$0.1<z<0.3$, and we have drawn interpolating lines. \ We note that since
$\ln(b_{N})$ and $\ln(b_{N-1})$ have approximately the same slope, most of the
$z$ dependence cancels in the combination $\ln(b_{N})-\ln(b_{N-1})$. \ This
suggests that $f\left(  z\right)  $ as defined in (\ref{fz}) is small. \ Much
of the systematic cutoff errors can be cancelled in the ratio $b_{N}/b_{N-1}%
$by keeping $z$ the same for $b_{N}$ and $b_{N-1}$. \ From Fig. \ref{z} we see
that $b_{N}/b_{N-1}$ is about $10$ for $5\leq N\leq10$. \ Therefore scaling
$B_{2}$ proportional to $10^{-N}$ as we probe the $N$-body droplet should keep
$z$ approximately the same for these values of $N$.%

\begin{figure}
[ptb]
\begin{center}
\includegraphics[
height=3.5267in,
width=3.4489in
]%
{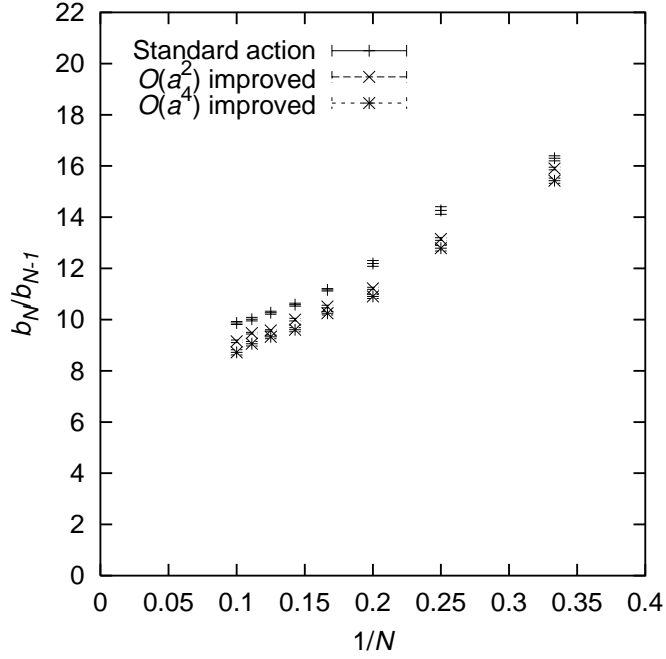}%
\caption{Comparsion of $b_{N}/b_{N-1}$ for the standard, $O(a^{2})$-improved,
and $O(a^{4})$-improved actions. \ We use $\varepsilon=3.7$ and $z_{2}%
=1.2\times10^{1-N}$.}%
\label{action}%
\end{center}
\end{figure}

Next we calculated $b_{N}/b_{N-1}$ for $N\leq10$ using three different
actions. \ We compared the standard action, the $O(a^{2})$-improved action,
and the $O(a^{4})$-improved action, using $a_{t}=(20$ MeV$)^{-1}$ and
$B_{2}=2\times10^{2-N}$ MeV. \ This corresponds with $\varepsilon=3.7$ and
$z_{2}=1.2\times10^{1-N}$. \ The results are shown in Fig. \ref{action}. \ We
see about a $10\%$ variation among the three different actions, with the
$O(a^{2})$- and $O(a^{4})$-improved actions agreeing slightly better with each
other than with the standard action.%

\begin{figure}
[ptb]
\begin{center}
\includegraphics[
height=3.5267in,
width=3.4489in
]%
{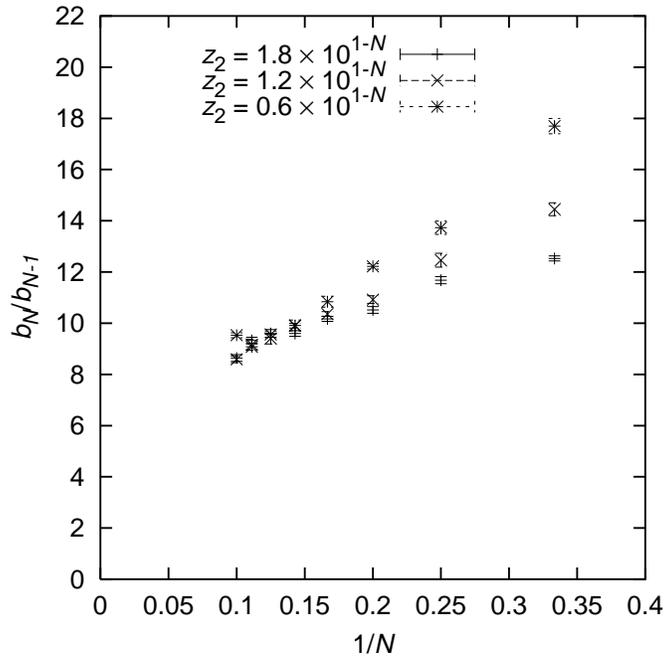}%
\caption{Comparsion of $b_{N}/b_{N-1}$ for different values of $z_{2}$. We use
the $O(a^{4})$-improved action and $\varepsilon=3.7$.}%
\label{b}%
\end{center}
\end{figure}

In Fig. \ref{b} we plot $b_{N}/b_{N-1}$ using the $O(a^{4})$-improved action,
$a_{t}=(20$ MeV$)^{-1}$, and three different sets of values for $B_{2}$:
$\ B_{2}=3\times10^{2-N}$ MeV, $2\times10^{2-N}$ MeV, and $1\times10^{2-N}$
MeV. \ This corresponds with $\varepsilon=3.7$ and $z_{2}=1.8\times10^{1-N}$,
$1.2\times10^{1-N}$, and $0.6\times10^{1-N}$ respectively.%
The discrepancies for the different values of $B_{2}$ are at the $30\%$ level
for small $N$, but as expected the errors decrease with increasing $N$.

\begin{figure}
[ptb]
\begin{center}
\includegraphics[
height=3.5267in,
width=3.4489in
]%
{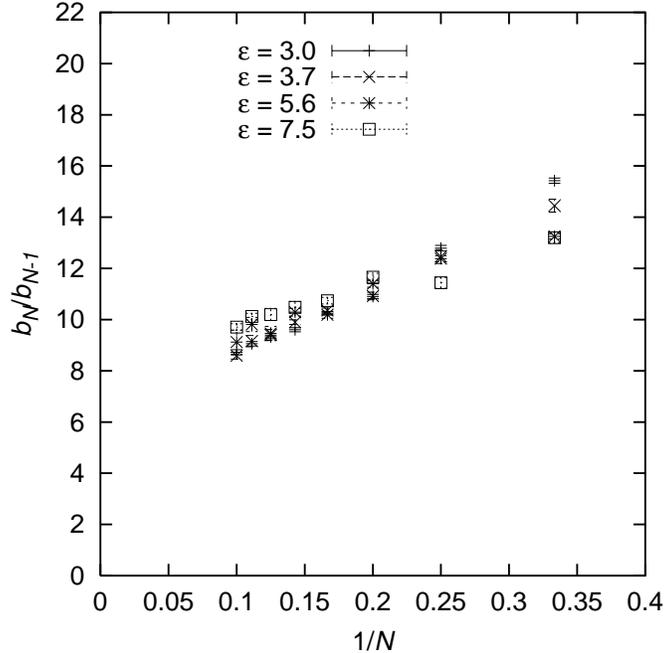}%
\caption{Comparsion of $b_{N}/b_{N-1}$ for different values of $\varepsilon$.
\ We use the $O(a^{4})$-improved action and $z_{2}=1.2\times10^{1-N}$.}%
\label{at}%
\end{center}
\end{figure}

We also studied the dependence of $b_{N}/b_{N-1}$ on the temporal lattice
spacing $a_{t}$. \ We set $B_{2}=2\times10^{2-N}$ MeV and used the $O(a^{4}%
)$-improved action with $a_{t}=(16$\ MeV$)^{-1}$, $(20$\ MeV$)^{-1}$,
$(30$\ MeV$)^{-1}$, and $(40$\ MeV$)^{-1}$. \ This corresponds with
$z_{2}=1.2\times10^{1-N}$ and $\varepsilon=3.0$, $3.7$, $5.6$, and $7.5$
respectively. Since $\varepsilon$ is rather large, $a_{t}$ has only a small
effect on the ultraviolet regularization of the two-body interaction.
\ Instead the importance of $a_{t}$ is as an auxiliary regulator on the
implicit $N$-body contact interaction. \ The results are shown in Fig.
\ref{at}.%
The results appear to differ at about the $10-15\%$ level.

\begin{figure}
[ptb]
\begin{center}
\includegraphics[
height=3.5276in,
width=3.4489in
]%
{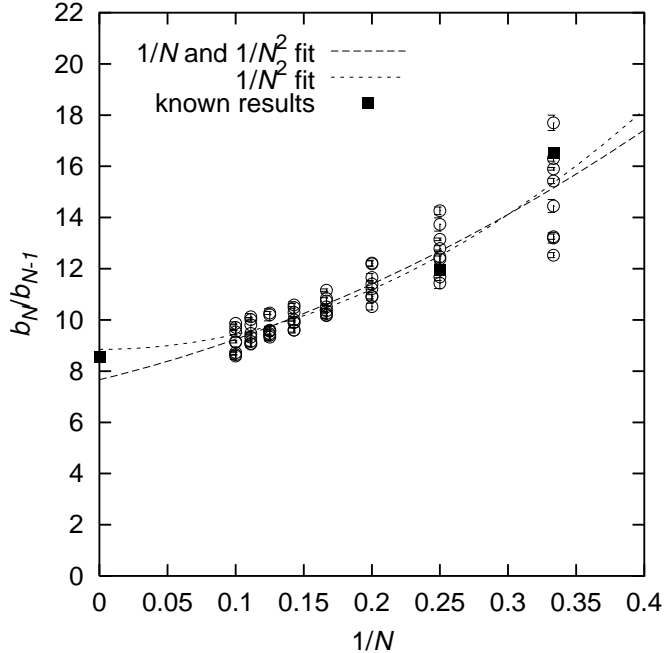}%
\caption{Cumulative data for $b_{N}/b_{N-1}$. \ We draw two best fit curves
with up to quadratic dependence on $1/N$ and show the known results for $N=3,$
$4,$ and $\infty$. }%
\label{all}%
\end{center}
\end{figure}

In Fig. \ref{all} we combine all of the data shown in Figs. \ref{action},
\ref{b}, and \ref{at}. \ For comparison we include the known results for $N=3$
\cite{Hammer:2004x}$,$ $N=4$ \cite{Platter:2004x}$,$ and $N\rightarrow\infty$
\cite{Hammer:2004x}. \ We draw two best fit curves with up to quadratic
dependence on $1/N$. \ The known results were not included in this fit. \ The
best fit curve using $1/N$ and $1/N^{2}$ gives a value%
\begin{equation}
\lim_{N\rightarrow\infty}\frac{b_{N}}{b_{N-1}}\simeq7.7,
\end{equation}
while the best fit curve using only $1/N^{2}$ gives a value%
\begin{equation}
\lim_{N\rightarrow\infty}\frac{b_{N}}{b_{N-1}}\simeq8.8.
\end{equation}
If we take these two results as approximate lower and upper bounds then we
find%
\begin{equation}
\lim_{N\rightarrow\infty}\frac{b_{N}}{b_{N-1}}\simeq8.3(6).
\end{equation}%

\section{Conclusions}

We have studied the two-dimensional $N$-particle system with short range
attraction using lowest-order lattice effective field theory. \ We discussed
two aspects of the large-$N$ limit. \ The first is a technique that uses
large-$N$ similarity under rescaling to cancel some of the nonzero range
corrections from the ratio $B_{N}/B_{N-1}$. \ The other is the problem of a
large implicit $N$-body contact interaction when many particles lie within a
region the size of the range of the potential. \ We regulated this implicit
$N$-body contact interaction on the lattice using a discrete
Hubbard-Stratonovich transformation. \ Using a heat bath/Metropolis algorithm
we computed $B_{N}/B_{N-1}$ for $N\leq10$. \ Extrapolating to the large-$N$
limit we found%
\begin{equation}
\lim_{N\rightarrow\infty}\frac{B_{N}}{B_{N-1}}=8.3(6)\text{.}%
\end{equation}
This appears to be in agreement with the value $8.567$ found by
\cite{Hammer:2004x}$.$

While we have measured the large-$N$ limit of $B_{N}/B_{N-1}$ to within
$10\%$, we relied on large-$N$ similarity under rescaling to keep the finite
cutoff errors in check. \ The $z$ dependence in Fig. \ref{z} suggests that one
needs to go beyond leading order to accurately describe all of the physics at
large $N$. \ This competition between effective field theory expansions and
the large-$N$ limit presents an interesting theoretical challenge. \ Since
there are no known physical systems where we can experimentally measure the
universal zero range behavior, the coefficients of the higher-dimensional
operators must be set by numerical calculations. \ One technique perhaps is to
use numerical renormalization group matching to relate the coefficients of
higher-dimensional operators for different values of $mB_{2}\Lambda^{-2}%
$.\ \ However more study would be needed to see if this is a viable technique.

Acknowledgments: The author is grateful to Hans-Werner Hammer and Lucas
Platter for discussions and for suggesting the problem. \ The author also
thanks Thomas Sch\"{a}fer for helpful discussions. \ This work is supported by
the US Department of Energy grant DE-FG02-04ER41335.

\bibliographystyle{apsrev}
\bibliography{NuclearMatter}

\end{document}